\documentclass[12pt]{iopart}
\usepackage{iopams}
\usepackage{bm}
\usepackage{graphicx}
\usepackage{epsfig}
\begin{document}

\title[Diluted ferromagnetic semiconductors]{Origin of ferromagnetic response in diluted magnetic semiconductors and oxides}

\author{Tomasz Dietl}

\address{Institute of Physics,
  Polish Academy of Sciences and ERATO Semiconductor Spintronics
  Project, al.~Lotnik\'ow 32/46, PL 02-668 Warszawa, Poland \\
Institute of Theoretical Physics, Warsaw University, ul. Ho\.za 69, 00-681 Warszawa, Poland}
\ead{dietl@ifpan.edu.pl}
\begin{abstract}
This paper reviews the present understanding of the origin of ferromagnetic response that has been detected in a number of diluted magnetic semiconductors (DMS) and diluted magnetic oxides (DMO) as well as in some nominally magnetically undoped materials. It is argued that these systems can be grouped into four classes. To the first belong composite materials in which precipitations of a known ferromagnetic, ferrimagnetic or antiferromagnetic compound account for magnetic characteristics at high temperatures. The second class forms alloys showing chemical nano-scale phase separation into the regions with small and large concentrations of the magnetic constituent. Here, high-temperature magnetic properties are determined by the regions with high magnetic ion concentrations, whose crystal structure is imposed by the host. Novel methods enabling a control of this spinodal decomposition and possible functionalities of these systems are described. To the third class belong (Ga,Mn)As, heavily doped p-(Zn,Mn)Te, and related semiconductors. In these solid solutions the theory built on p-d Zener's model of hole-mediated ferromagnetism and on either the Kohn-Luttinger $kp$ theory or the multi-orbital tight-binding approach describes qualitatively, and often quantitatively, thermodynamic, micromagnetic, optical, and transport properties. Moreover, the understanding of these materials has provided a basis for the development of novel methods enabling magnetisation manipulation and switching. Finally, in a number of carrier-doped DMS and DMO a competition between long-range ferromagnetic and short-range antiferromagnetic interactions and/or the proximity of the localisation boundary lead to an electronic nano-scale phase separation.  These materials exhibit characteristics similar to colossal magnetoresistance oxides.

\end{abstract}

\maketitle

\section{Introduction}

Since a number of years a considerable attention has been devoted to materials systems combining properties and functionalities specific to both semiconductors and ferromagnets. For instance, stray magnetic fields in hybrid heterostructures consisting of semiconductor and ferromagnetic metallic layers \cite{Prinz:1998_a}  generate spin currents {\em via} Zeeman and Stern-Gerlach effects \cite{Dietl:2006_b}. Furthermore, spin-polarised electrons in the metal can be injected into or across the semiconductor \cite{Hanbicki:2003_a}. The development of such spin injectors brings us closer to the realisation of spin transistors and stable surface emitting lasers \cite{Rudolph:2003_a}. At the same time, spin-dependent boundary conditions and/or stray fields of the ferromagnet can affect magnetotransport and magnetooptical properties of the semiconductor, a phenomenon exploited in novel optical isolators \cite{Shimizu:2002_a,Shimizu:2004_a}. Particularly perspective materials in the context of hybrid structures appear to be those elemental or compound ferromagnets which can be grown in the same reactor as the semiconductor counterpart.

At the same time, an enormous activity has been directed to the development of functional magnetic semiconductors. Here, two approaches have been pursued. First of them, initiated by Ga{\l}{\c{a}}zka {\it et al.}  \cite{Galazka:1978_a,Story:1986_a,Dietl:1994_a} and by Ohno {\it et al.} \cite{Ohno:1992_a,Ohno:1996_a,Matsukura:2002_c}, is the incorporation of magnetism to functional semiconductors by alloying with transition metals, notably Mn.  By now, a ferromagnetic response, often persisting up to above room temperature, has been detected in a large number of semiconductor and oxide thin layers containing a minute amount of magnetic ions \cite{Dietl:2005_c,Liu:2005_a} or even nominally undoped by magnetic elements \cite{Coey:2005_a}. The other relevant approach is the improvement of semiconductor functionalities of the existing magnetic oxides, such as ferrites and garnets, by exploiting the recent progress in oxide epitaxy and modifications \cite{Dumont:2005_a,Luders:2006_a}.

In this paper, we first outline various experimental difficulties in the unambiguous determination of the origin of magnetic signals in the case of thin diluted films, emphasising the existence of many reasons of the signal contamination by magnetic nanoparticles.  We then address the question about the possibility of high temperature ferromagnetic ordering in materials containing no magnetic elements. The main body of the paper is devoted to possible origins of ferromagnetic response in diluted magnetic semiconductors (DMS) and diluted magnetic oxides (DMO), in which {\em no} precipitates of foreign compound occur.  Therefore, the ferromagnetism of such DMS and DMO does not result from the presence of any known ferromagnetic, ferrimagnetic or antiferromagnetic nanoparticles. We distinguish three groups of crystallographically uniform DMS and DMO.

First of them form ferromagnetic DMS, such as (Ga,Mn)As as well as heavily doped p-(Zn,Mn)Te and p-(Pb,Mn)Te, in which long-range exchange interactions between Mn spins are mediated by holes in the valence band. However, these ferromagnetic systems by no means exhaust all possibilities. In particular, a question  arises whether at the length scale appropriately greater that the mean distance between magnetic ions a spatially uniform ferromagnetic spin order is a real ground state of all DMS and DMO, in which ferromagnetic signatures have been discovered.  Actually, the existence and the role of spatially {\em non}-uniform ferromagnetic spin order has been an important theme in research on both magnetic semiconductors \cite{Nagaev:1993_a} and colossal magnetoresistance oxides \cite{Dagotto:2001_a}. Nano-scale phase separation effects that were invoked to explained pertinent properties of those materials may {\em a priori} be even more relevant in ferromagnetic DMS and DMO, in which carrier correlation and electrostatic disorder associated with ionised impurities co-exist with chemical disorder in the cation sublattice. In this paper, we emphasise the importance of two kinds of nano-scale spacial phase separations relevant to DMS and DMO. First is spinodal alloy decomposition -- {\em chemical} phase separation -- into regions incorporating either very large and very small concentration of the magnetic constituent. Second is disorder-driven {\em electronic} phase separation into ferromagnetic bubbles containing a large carrier density, and immersed in a depleted paramagnetic environment.

Two main conclusions can be drawn from the results presented here. First,  weakly localised or delocalised carriers are necessary to mediate efficient ferromagnetic interactions between randomly distributed diluted magnetic spins. Second, there is a growing amount of evidences that the distribution of magnetic ions over cation sites can be controlled in a number of DMS and DMO by growth conditions, co-doping, and post-growth processing. This makes it possible to prepare a given material with properties and functionalities specific to either uniform magnetic semiconductors or hybrid semiconductor/ferromagnet composite systems.

\section{Effects of magnetic nanoparticles}

As known, a highly sensitive SQUID magnetometer is necessary to detect small signals produced by thin films only weakly doped by magnetic elements. Actually,  the magnetic moment of the film is often inferior to that coming from typical sample holders, substrates, magnet remanence, or that produced by non-ideal signal processing software. Furthermore, the film magnetic response can be contaminated by the presence of magnetic nanoparticles. Such nanoparticles may originate from residual impurities in the growth chamber or source materials as well as can be introduced by post-growth processing procedures \cite{Abraham:2005_a}.  Furthermore, in view of limited solubility of magnetic atoms in semiconductors, and the existence of many magnetic compounds involving elements in question, nanoparticles containing a large concentration of magnetic ions can precipitate during the growth or processing.

Obviously, both ferromagnetic and ferrimagnetic nanoparticles possess a non-zero magnetic moment up to the ordering temperature $T_{\mathrm{m}}$ of the corresponding compound, typically up to well above the room temperature. Interestingly enough, nanoparticles in which antiferromagnetic interactions dominate can also show a non-zero magnetic moment due to the presence of uncompensated  spins at their surface, whose relative magnitude grows with decreasing nanoparticle size \cite{Trohidou:2002_a}. Importantly, even uncoupled nano-scale magnetised regions give rise to macroscopic ferromagnetic signatures, such as spontaneous magnetisation and magnetic hysteresis, up to the blocking temperature $T_{\mathrm{B}} \leq T_{\mathrm{m}}$, where (e.~g.~\cite{Shinde:2004_a}) $T_{\mathrm{B}} = KV/[k_{\mathrm{B}}\ln(t_{\mathrm{lab}}/\tau)]$. Here $K$ is the density of the magnetic anisotropy energy, $V$ is nanoparticle volume, and $\ln(t_{\mathrm{lab}}/\tau) \approx 25$ for a typical ratio of a relevant spin-flip relaxation time $\tau$ to the time of hysteresis measurements, $t_{\mathrm{lab}}$. The value of $T_{\mathrm{B}}$ is further increased by dipolar interactions between nanoparticles, which deteriorate also the squareness of the hysteresis loop.

\section{Is ferromagnetism possible in semiconductors with no magnetic elements?}

Organic ferromagnets \cite{Fujiwara:2005_a} and quantum Hall ferromagnets \cite{Jungwirth:2000_a} demonstrate that ferromagnetism is possible in materials without magnetic ions, albeit the corresponding Curie temperatures are rather low, below 20~K. It has also been suggested that a robust ferromagnetism can appear in certain zinc-blende metals, such as CaAs, driven by a Stoner instability in the narrow heavy holes band \cite{Geshi:2005_a}. There is, however, a number of qualitative indications against the persistence of ferromagnetism up to above room temperature in semiconductors, oxides, or carbon derivatives incorporating no magnetic elements and, at the same time, containing only a low carrier density, as witnessed by the high resistivity.

It has been known for a long time that a number of defects or non-magnetic impurities form localised paramagnetic centers in non-metals. However, it is not easy to find a mechanism which would generate a ferromagnetic interaction between such spins, which -- in view of high $T_{\mathrm{C}}$ in question -- has to be rather strong. Furthermore, it should be long range, as the concentration of the relevant centers is low, according to a small magnitude of saturation magnetisation. A large distance between the spins is in fact required by the Mott criterion, according to which an average distance between the localised centers is at least 2.5 times larger than their localisation radius in non-metals.  Moreover, in the absence of carriers, exchange interactions between spins are typically not only short-range but also merely antiferromagnetic \cite{Bhatt:1982_a}. Only in rather few cases, a compensation of various exchange contributions leads to net ferromagnetic interactions between localised spins. Celebrated examples are actually ferromagnetic semiconductors, such EuO and CdCr$_2$Se$_4$, in which despite the absence of carriers the coupling between highly localised f and d spins is ferromagnetic. However, the Curie temperature barely exceeds 100~K in these materials though the spin concentration is rather high.

In view of the facts above it is not surprising that it is hard to invent a sound model explaining high Curie temperatures in materials in which, say, 5\% of sites contains a spin, and only short-range exchange interactions can operate. At this stage it appears more natural to assume that a small number of magnetic nanoparticles which escaped from the detection procedure account for the ferromagnetic-like behaviour of nominally nonmagnetic insulators.

Finally, we note that a spontaneous magnetic moment reported for the case of some nonmagnetic nanostructures has been assigned to orbital magnetism, enhanced by a spin-orbit interaction \cite{Hernando:2006_a}.

\section{Nonuniform ferromagnetic DMS -- chemical nano-scale phase separations}
\subsection{Spinodal decomposition}

It is well known that phase diagrams of a number of alloys exhibit a solubility gap in a certain concentration range. This may lead to a spinodal decomposition into regions with a low and a high concentration of particular constituents, a mechanism suggested to account for high Curie temperatures for a number of DMS \cite{Dietl:2005_c}. If the concentration of one of them is small, it may appear in a form of coherent nanocrystals embedded by the majority component. For instance, such a spinodal decomposition is known to occur in the case of (Ga,In)N \cite{Farhat:2002_a}, where In rich quantum-dot like regions are embedded by an In poor matrix. However, according to the pioneering {\em ab initio} work of van Schilfgaarde and Mryasov \cite{Schilfgaarde:2001_a} and others \cite{Sato:2005_a} particularly strong tendency to form non-random alloy occurs in the case of DMS: the evaluated gain in energy by bringing two Ga-substitutional Mn atoms together is $E_{\mathrm{d}} = 120$~meV in GaAs and 300~meV in GaN, and reaches 350~meV in the case of Cr pair in GaN \cite{Schilfgaarde:2001_a}. We would like to emphasise that magnetic nanostructures embedded by the semiconductor or oxide in question may assume a novel crystallographic form and/or chemical composition, unclassified so-far in materials compendia.

Since spinodal decomposition does not usually involve a precipitation of another crystallographic phase, it is not easy detectable experimentally. Nevertheless, its presence was found by electron transmission microscopy (TEM) \cite{Moreno:2002_a,Yokoyama:2005_a} in (Ga,Mn)As, where coherent zinc-blende Mn-rich (Mn,Ga)As metallic nanocrystals led to the apparent Curie temperature up to 360~K \cite{Yokoyama:2005_a}, as shown in Fig.~1. Furthermore, coherent hexagonal and diamond-type Mn-rich nanocrystals were detected by spatially resolved x-ray diffraction in (Ga,Mn)N \cite{Martinez-Criado:2005_a} and by transmission electron microscopy in (Ge,Mn) \cite{Ahlers:2006_a,Jamet:2006_a}, respectively.  In view of typically rather limited solubility of magnetic elements in semiconductors, it may be expected that such a spinodal decomposition is a generic property of a number of DMS and DMO. The nanocrystals form in this way may only be stable inside the semiconductor matrix. Accordingly, whether they will exhibit ferromagnetic, ferrimagnetic or antiferromagnetic order is {\em a priori}  unknown. However, owing to the high concentration of the magnetic constituent, the ordering temperature $T_m$ is expected to be relatively high, typically above the room temperature. In addition to spontaneous magnetisation expected for ferromagnetic and ferromagnetic nanoparticles, we recall that uncompensated spins at the surface of antiferromagnetic nanocrystals can also result in a sizable value of spontaneous magnetisation below, often high, N\'eel temperature \cite{Trohidou:2002_a}.

\begin{figure}
\includegraphics[scale=0.7]{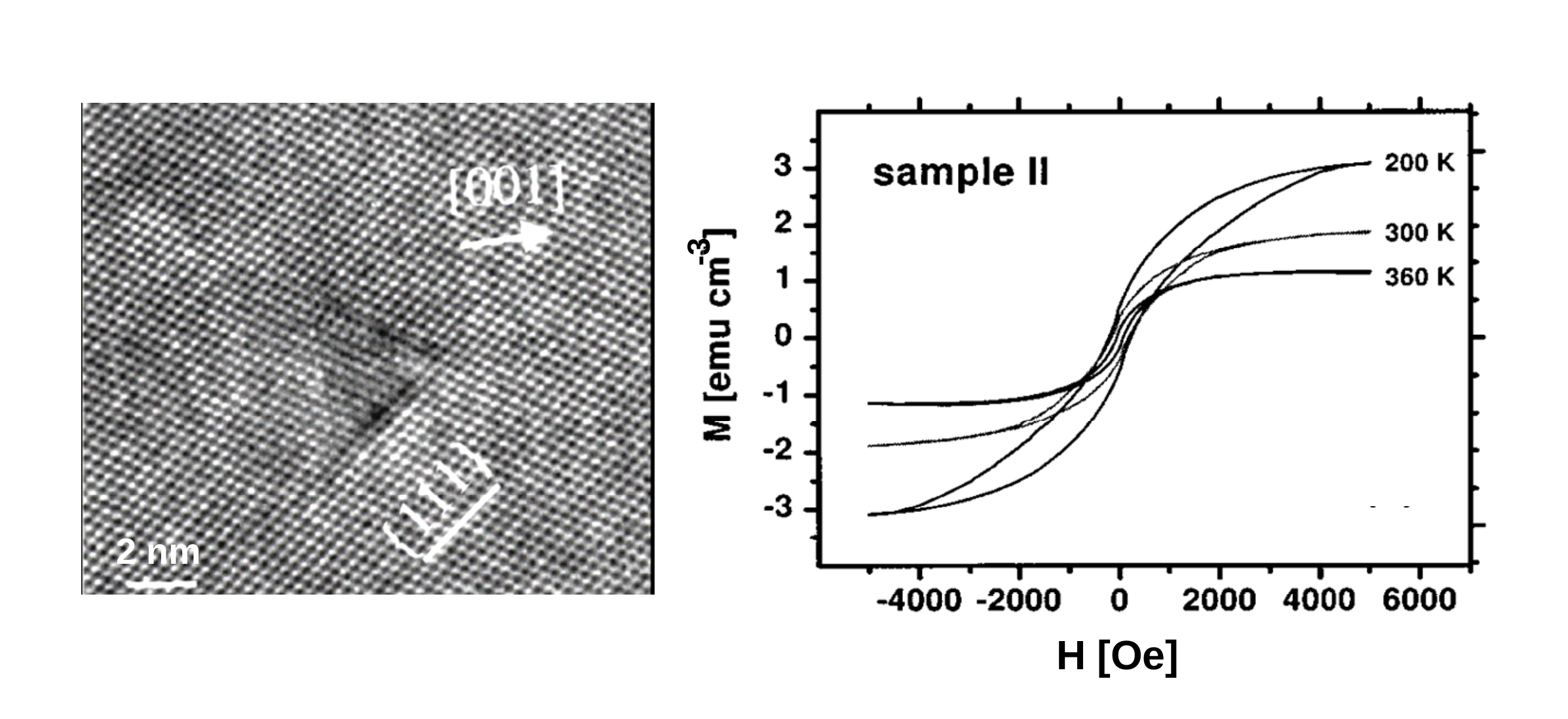}
\caption{Transmission electron micrograph of zinc-blende Mn-rich (Mn,Ga)As nanocrystal in GaAs (left panel) and the corresponding magnetisation hysteresis at various temperatures (right panel). After \cite{Moreno:2002_a}.}
\end{figure}

As an example we consider (Zn,Cr)Se \cite{Karczewski:2003_a} and (Zn,Cr)Te \cite{Saito:2003_a}, which show \cite{Karczewski:2003_a,Kuroda:2005_a} the well-known superparamagnetic behaviour \cite{Shinde:2004_a,Goswami:2005_a}, indicating that the system is to be viewed rather as an ensemble of non-interacting ferromagnetic particles than a uniform magnetic alloy. In such a case the temperature dependencies of magnetisation and magnetic susceptibility are described by four distinguished temperatures: $T_m$, the blocking temperature $T_{\mathrm{B}}$ that corresponds to a maximum of zero-field cooled magnetisation; the apparent Curie temperature $T_{\mathrm{C}}^{\mathrm{(app)}}$ of the composite material, and the Curie-Weiss temperature $\Theta$ characterizing a weighted magnitude of the exchange interactions between the Cr spins within the nanocrystal. A maximum value of $T_{\mathrm{C}}^{\mathrm{(app)}} \approx 320$~K is obtained for (Zn,Cr)Te with $x_{\mathrm{Cr}}$ above the percolation limit for 3D, $x \approx 0.2$.

These remarkable observations can readily be interpreted under the assumption that the relevant magnetic nanoparticles are built of a metallic zinc-blende CrTe or Cr-rich (Zn,Cr)Te characterised by $T_m  \approx 320$~K and by the lattice constant imposed by a paramagnetic semiconductor host, either ZnTe or (Zn,Cr)Te with a rather small Cr concentration. This conjecture is consistent with {\em ab initio} computations \cite{Zhao:2005_a} predicting zinc-blende CrTe to be a ferromagnetic half-metal as well as with experimental results for CrTe in a bulk NiAs-type structure for which $T_{\mathrm{m}} \equiv T_{\mathrm{C}} = 340 \pm 10$~K \cite{Ohta:1993_a}. Within this scenario, for small ferromagnetic nanocrystals we expect $T_{\mathrm{B}} < T_{\mathrm{C}}^{\mathrm{(app)}} < T_{\mathrm{C}}$, $T_{\mathrm{B}}$ being proportional to a mean nanocrystal volume $V$, $T_{\mathrm{B}} \approx KV/(25k_{\mathrm{B}})$, where $K$ is the density of the magnetic anisotropy energy. Similarly, $T_{\mathrm{C}}^{\mathrm{(app)}}$ provides information on the upper bound of the $V$ distribution. Furthermore, we note that broken magnetic bonds at the nanocrystal surface reduce the Curie-Weiss temperature $\Theta$ from its anticipated value for large $V$, $\Theta_{\mathrm{max}} \geq T_{\mathrm{C}}$.

It is, therefore, legitimate to suppose that coherent nanocrystals with a large concentration of magnetic constituent account for high apparent Curie temperatures detected in a number of DMS and DMO. This model explains, in particular, a long staying puzzle about the origin of ferromagnetic response in DMS and DMO, in which an average concentration of magnetic ions is below the percolation limit for the nearest neighbour coupling and, at the same time, the free carrier density is too low to mediate an efficient long-range exchange interaction. Remarkably, the presence of magnetically active metallic nanocrystals leads to enhanced magnetotransport \cite{Shinde:2004_a,Ye:2003_a} and magnetooptical \cite{Yokoyama:2005_a} properties over a wide spectral range. This opens doors for various applications of such hybrid systems provided that methods for controlling nanocrystal characteristics and distribution would be elaborated. So far, the most rewarding method of self-organized growth of coherent nanocrystals or quantum dots has exploited strain fields generated by lattice mismatch at interfaces of heterostructures \cite{Stangl:2004_a}.  Remarkably, it becomes possible to fabricate highly ordered three dimensional dot crystals under suitable spatial strain anisotropy \cite{Stangl:2004_a}. A further progress in this direction is particularly timely as it could result in the development of high-density 3D memories and spatial light modulators for advanced photonic applications \cite{Park:2002_a}. A new  method of self-organized growth has recently been proposed by the present author \cite{Dietl:2006_a}. In this approach, long-range Coulomb forces serve to control the aggregation of alloy constituents.

\subsection{Controlling spinodal decomposition by inter-ion Coulomb interactions}

It is well know that in most DMS the levels derived from the open d or f shells of magnetic ions reside in the band gap of the host semiconductor \cite{Dietl:2002_d}. This property of magnetic ions has actually been exploited for a long time to fabricate semi-insulating materials, in which carriers introduced by residual impurities or defects are trapped by the band-gap levels of magnetic impurities. The essential ingredient of the proposal in question \cite{Dietl:2006_a} is the observation that such a trapping alters the charge state of the magnetic ions and, hence, affect their mutual Coulomb interactions. Accordingly, co-doping of DMS with shallow acceptors or donors, during either growth or post-growth processing, modifies $E_{\mathrm{d}}$ and thus provides a mean for the control of ion aggregation. Indeed, the energy of the Coulomb interaction between two elementary charges residing on the nearest neighbour cation sites in the GaAs lattice is 280~meV. This value indicates that the Coulomb interaction can preclude the aggregation, as the gain of energy associated with the bringing two Mn atoms in (Ga,Mn)As is $E_{\mathrm{d}} = 120$~meV.

It is evident that the model in question should apply to a broad class of DMS as well to semiconductors and insulators, in which a constituent, dopant, or defect can exist in different charge states under various growth conditions. As important examples we consider (Ga,Mn)N and (Zn,Cr)Te, in which remarkable changes in ferromagnetic characteristics on co-doping with shallow impurities have recently been reported \cite{Reed:2005_a,Ozaki:2005_a}. In particular, a strong dependence of saturation magnetisation $M_{\mathrm{s}}$ at 300~K on co-doping with Si donors and Mg acceptors has been found \cite{Reed:2005_a} for (Ga,Mn)N with an average Mn concentration $x_{\mathrm{Mn}} \approx 0.2$\%. Both double exchange and superexchange are inefficient at this low Mn concentration and for the mid-gap Mn level in question. At the same time, the model of nanocrystal self-organized growth in question explains readily why $M_{\mathrm{s}}$ goes through a maximum when Mn impurities are in the neutral Mn$^{\mathrm{3+}}$ state, and vanishes if co-doping by the shallow impurities makes all Mn atoms to be electrically charged.

It has also been found that $T_{\mathrm{C}}^{\mathrm{(app)}}$ depends dramatically on the concentration of shallow N acceptors in (Zn,Cr)Te, an effect shown in Fig.~2. Actually, $T_{\mathrm{C}}^{\mathrm{(app)}}$ decreases monotonically when the concentration $x_{\mathrm{N}}$ of nitrogen increases, and vanishes when $x_{\mathrm{Cr}}$ and $x_{\mathrm{N}}$ become comparable. This supports the model as in ZnTe the Cr state \cite{Godlewski:1980_a} resides about 1~eV above the nitrogen level \cite{Baron:1998_a}. Accordingly, for $x_{\mathrm{N}} \approx x_{\mathrm{Cr}}$ all Cr atoms become ionised and the Coulomb repulsion precludes the nanocrystal formation. At the same time, the findings are not consistent with the originally proposed double exchange mechanism \cite{Ozaki:2005_a}, as undoped ZnTe is only weakly p-type, so that $T_{\mathrm{C}}$ should be small for either $x_{\mathrm{N}} \approx 0$ and $x_{\mathrm{N}} \approx x_{\mathrm{Cr}}$, and pick at $x_{\mathrm{N}} \approx x_{\mathrm{Cr}}/2$, not at $x_{\mathrm{N}} \approx 0$.

\begin{figure}
\includegraphics[scale=1.8]{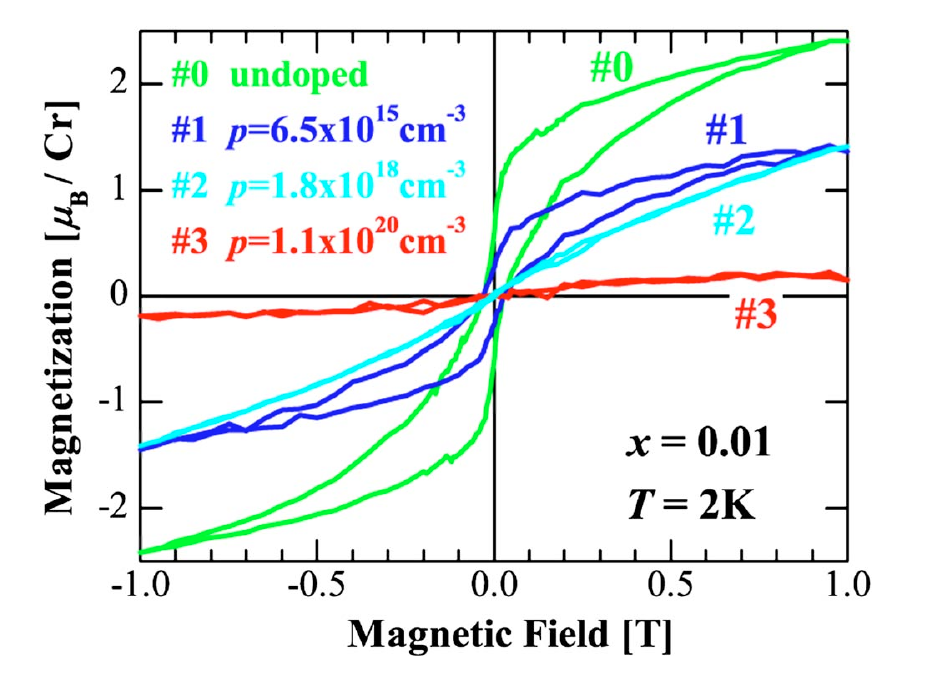}
\caption{Effect of doping by nitrogen acceptors on magnetisation of Zn$_{0.99}$Cr$_{0.01}$Te. After \cite{Ozaki:2005_a}.}
\end{figure}

Finally, we mention the case of Mn doped GaAs, InAs, GaSb, and InSb. In those materials, owing to a relatively shallow character of Mn acceptors and a large Bohr radius, the holes reside in the valence band. Thus, the Mn atoms are negatively charged, which - according to our model -- reduces their clustering, and makes it possible to deposit, by low-temperature epitaxy, a uniform alloy with a composition beyond the solubility limit. Co-doping with shallow donors, by reducing the free-carrier screening, will enhance repulsions among Mn, and allow one to fabricate homogenous layers with even greater $x_{\mathrm{Mn}}$. On the other hand, co-doping by shallow acceptors, along with the donor formation by a self-compensation mechanism \cite{Yu:2002_a}, will enforce the screening and, hence, lead to nanocrystal aggregation.

\section{Spatially uniform ferromagnetic DMS}

Since for decades III-V semiconductor compounds have been applied as photonic and microwave devices, the discovery of ferromagnetism first in (In,Mn)As \cite{Ohno:1992_a} and then in (Ga,Mn)As \cite{Ohno:1996_a} came as a landmark achievement. In these materials, substitutional divalent Mn ions provide localised spins and function as acceptor centers that provide holes which mediate the ferromagnetic coupling between the parent randomly distributed Mn spins \cite{Dietl:1997_a,Matsukura:1998_a,Jungwirth:1999_a}.  In another technologically important group of semiconductors, in II-VI compounds, the densities of spins and carriers can be controlled independently, similarly to the case of IV-VI materials, in which hole-mediated ferromagnetism was discovered already in the 1980s \cite{Story:1986_a}. Stimulated by the theoretical predictions \cite{Dietl:1997_a}, search for carrier-induced ferromagnetism in II-IV materials containing Mn was undertaken. Experimental studies conducted with the use of magnetooptical and magnetic methods led to the discovery of ferromagnetism in 2D \cite{Haury:1997_a} and 3D \cite{Ferrand:2001_a} II-VI Mn-based DMS doped by nitrogen acceptors.

Since magnetic properties are controlled by band holes, an appealing possibility is to influence the magnetic ordering isothermally, by light or by the electric field, which affect the carrier concentration in semiconductor quantum structures. Such tuning capabilities of the materials systems in question were put into the evidence in (In,Mn)As/(Al,Ga)Sb \cite{Koshihara:1997_a,Ohno:2000_a} and modulation doped p-(Cd,Mn)Te/(Cd,Mg,Zn)Te \cite{Haury:1997_a,Boukari:2002_a} heterostructures. Actually, these findings can be quantitatively interpreted by considering the effect of the electric field or illumination on the hole density under stationary conditions and, therefore, on the Curie temperature in the relevant magnetic layers. Interestingly, according to experimental findings and theoretical modelling, photocarriers generated in II-VI systems by above barrier illumination destroy ferromagnetic order in the magnetic quantum well residing in an undoped (intrinsic) region of a p-i-p structure \cite{Haury:1997_a,Boukari:2002_a} but they enhance the magnitude of spontaneous magnetisation  in the case of a p-i-n diode \cite{Boukari:2002_a}. Furthermore, the current-induced magnetisation  reversal was demonstrated in submicron pillars of (Ga,Mn)As/GaAs/(Ga,Mn)As \cite{Chiba:2004_b,Elsen:2006_a}.  Spin-polarised current was also shown to displace magnetic domain walls in (Ga,Mn)As with the easy axis perpendicular to the film plane \cite{Yamanouchi:2004_a,Yamanouchi:2006_a}.

Guided by the growing amount of experimental results, including informative magnetic resonance \cite{Szczytko:1999_b,Fedorych:2002_a} and photoemission \cite{Mizokawa:2002_a,Rader:2004_a,Hwang:2005_b} studies, a theoretical model of the hole-controlled ferromagnetism in III-V, II-VI, and group IV semiconductors containing Mn was proposed \cite{Dietl:2000_a,Dietl:2001_b}.  These materials exhibit characteristics specific to both charge transfer insulators and strongly correlated disordered metals. Moreover, complexities specific to strongly correlated systems coexist in DMS with features exhibited by heavily doped semiconductors and semiconductor alloys, such as Anderson-Mott localisation \cite{Dietl:1994_a}, defect generation by self-compensation mechanisms \cite{Dietl:2001_b,Masek:2001_a,Yu:2002_a}, and the breakdown of the virtual crystal approximation \cite{Benoit:1992_a}. Nevertheless,  the theory built on p-d Zener's model of carrier-mediated ferromagnetism and on either Kohn-Luttinger's $kp$ \cite{Dietl:2000_a,Dietl:2001_b,Abolfath:2001_a} or multi-orbital tight-binding \cite{Vurgaftman:2001_a,Sankowski:2005_a,Timm:2005_b} descriptions of the valence band in tetrahedrally coordinated semiconductors has qualitatively, and often quantitatively, described thermodynamic, micromagnetic, transport, and optical properties of DMS with delocalised holes \cite{Dietl:2004_a,Jungwirth:2006_a,Sankowski:2006_b}, challenging competing theories. It is often argued that owing to these studies (Ga,Mn)As has become one of the best-understood ferromagnets. Accordingly, this material is now serving as a testing ground for various \emph{ab initio} computation approaches to strongly correlated and disordered systems.

It should be emphasized that the  description of ferromagnetic DMS in terms of the above mentioned p-d Zener model is strictly valid only in the weak coupling limit \cite{Dietl:2000_a}. On going from antimonides to nitrides or from tellurides to oxides, the p-d hybridization increases. In the strong-coupling limit, the short-range part of the TM potential may render the virtual-crystal approximation and molecular-field approximation invalid \cite{Benoit:1992_a,Dietl:2002_c}, leading to properties somewhat reminiscent of those specific to non-VCA alloys, like Ga(As,N) \cite{Wu:2002_a}. In particular, the short-range potential leads to a local admixture to extra atomic wave functions, which can produce modifications in optical and transport characteristics, such as an apparent change in carrier effective masses. The issue how various corrections to the mean-field p-d Zener model \cite{Dietl:2000_a} affect theoretical values of $T_{\mathrm{C}}$ was recently examined in some detail for (Ga,Mn)As \cite{Timm:2005_b,Jungwirth:2005_b} with the conclusions that the overall picture remains quantitatively valid. Figure 3 shows one of the recent theoretical modelings of $T_{\mathrm{C}}$ in comparison to experimental findings for (Ga,Mn)As \cite{Jungwirth:2005_b}. These results confirm, in particular, that $T_{\mathrm{C}}$ values above 300~K could be achieved in Ga$_{0.9}$Mn$_{0.1}$As if such a large magnitude of the substitutional Mn concentration could be accompanied by a corresponding increase of the hole density \cite{Dietl:2000_a}.

\begin{figure}
\includegraphics{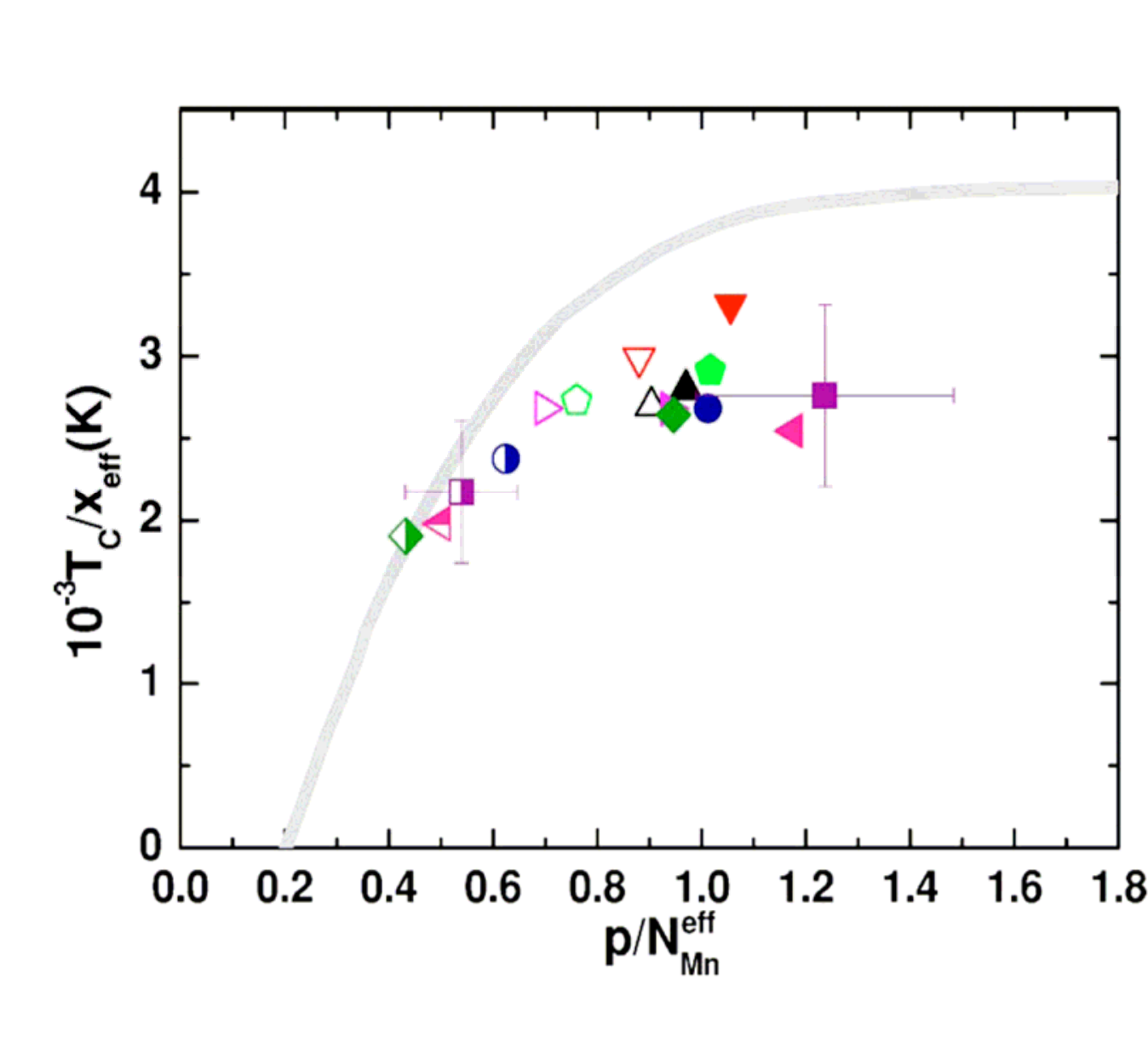}
\caption{Experimental Curie temperatures {\em vs.} hole density $p$ relative to
effective concentration of Mn moments, $N_{Mn} = 4x_{eff}/a_o^3$, where
$x_{eff}$ is the content of Mn in the substitutional positions and $a_o$ is the
lattice constant.  Grey line is theoretical computed within
the tight binding and coherent potential approximations. After \cite{Jungwirth:2005_b}.}
\end{figure}

\section{Nonuniform ferromagnetic DMS -- electronic nano-scale phase separations}
\subsection{Effects of competing magnetic interactions}

A number of effects has been identified, which may lead to deviations from a simple ferromagnetic spin order in carrier-controlled diluted ferromagnetic semiconductors even if the spatial distribution of magnetic ions is uniform. In particular, spin-density waves appear to be the ground state in the case of 1D systems \cite{Dietl:1999_a}. Another proposal involves canted ferromagnetism stemming from  a non-scalar form of spin-spin interactions, brought about by spin-orbit coupling \cite{Zarand:2002_a}, though a large value of saturation magnetisation in (Ga,Mn)As indicates that the effect is not large \cite{Jungwirth:2006_b}. Finally, a competition between long-range ferromagnetic interactions and intrinsic short-range antiferromagnetic interactions \cite{Kepa:2003_a} may affect the character of magnetic order \cite{Kechrakos:2005_a}. It appears that the effect is more relevant in II-VI DMS  than in III-V DMS where Mn centers are ionised, so that the enhanced hole density at closely lying Mn pairs may compensate antiferromagnetic interactions \cite{Dietl:2001_b}.

The above mentioned competition between the long-range RKKY merely ferromagnetic interaction and short-range merely antiferromagnetic superexchange was shown to affect in a nontrivial way magnetic properties of modulation-doped p-type (Cd,Mn)Te quantum wells \cite{Kechrakos:2005_a}. In this system, the temperature $T_{\mathrm{C}}$ at which spontaneous spin splitting of electronic levels appears as well as its temperature dependence \cite{Haury:1997_a,Kossacki:2000_a,Boukari:2002_a,Kossacki:2002_a} follow predictions of a simple mean-field Zener-like model of ferromagnetism \cite{Dietl:1997_a}. A reasonable accuracy of the mean-field approximation in this low-dimensional system was linked to the long-range character of the ferromagnetic interactions as well as to the combined effects of spin-orbit interaction and confinement that lead to the Ising-type universality class \cite{Haury:1997_a}. At the same time, however, wide hysteresis loops and the associated spontaneous macroscopic magnetisation in zero magnetic field, which are expected within this model \cite{Lee:2002_b}, have not been observed. Instead, according to polarisation-resolved photoluminescence measurements, the global spin polarisation of the carrier liquid increases slowly with the external magnetic field along the easy axis, reaching saturation at a field by a factor of twenty greater than what could be accounted for by demagnetisation effects \cite{Kossacki:2000_a,Kossacki:2002_a}.

In order to explain this behaviour, Monte Carlo simulations were employed \cite{Kechrakos:2005_a}, in which the Schr\"odinger equation was solved at each Monte Carlo sweep. Such a model is capable to assess the influence of magnetisation fluctuations, short-range antiferromagnetic interactions, disorder, magnetic polaron formation, and spin-Peierls instability on the carrier-mediated ferromagnetism in two-dimensional electronic systems. It has been found that the carriers generate ferromagnetic ordering but both critical temperature and hysteresis are affected in a nontrivial way by the presence of short-range antiferromagnetic interactions, as shown in Fig.~4. In particular, the antiferromagnetic interactions decrease $T_{\mathrm{C}}$ less than expected within the mean-field approximation. However, the presence of competing interactions reduces strongly the remanence and the coercive field. It appears that in order to satisfy both ferromagnetic and antiferromagnetic spin couplings in an optimum way, the system breaks into nano-scale ferromagnetic domains, so that the global magnetisation averages to zero. At the same time, the magnitude of the external magnetic field that can align these domains is set by the strength of the antiferromagnetic interactions, not by demagnetisation or magnetic anisotropy energies.

\begin{figure}
\includegraphics[scale=0.6]{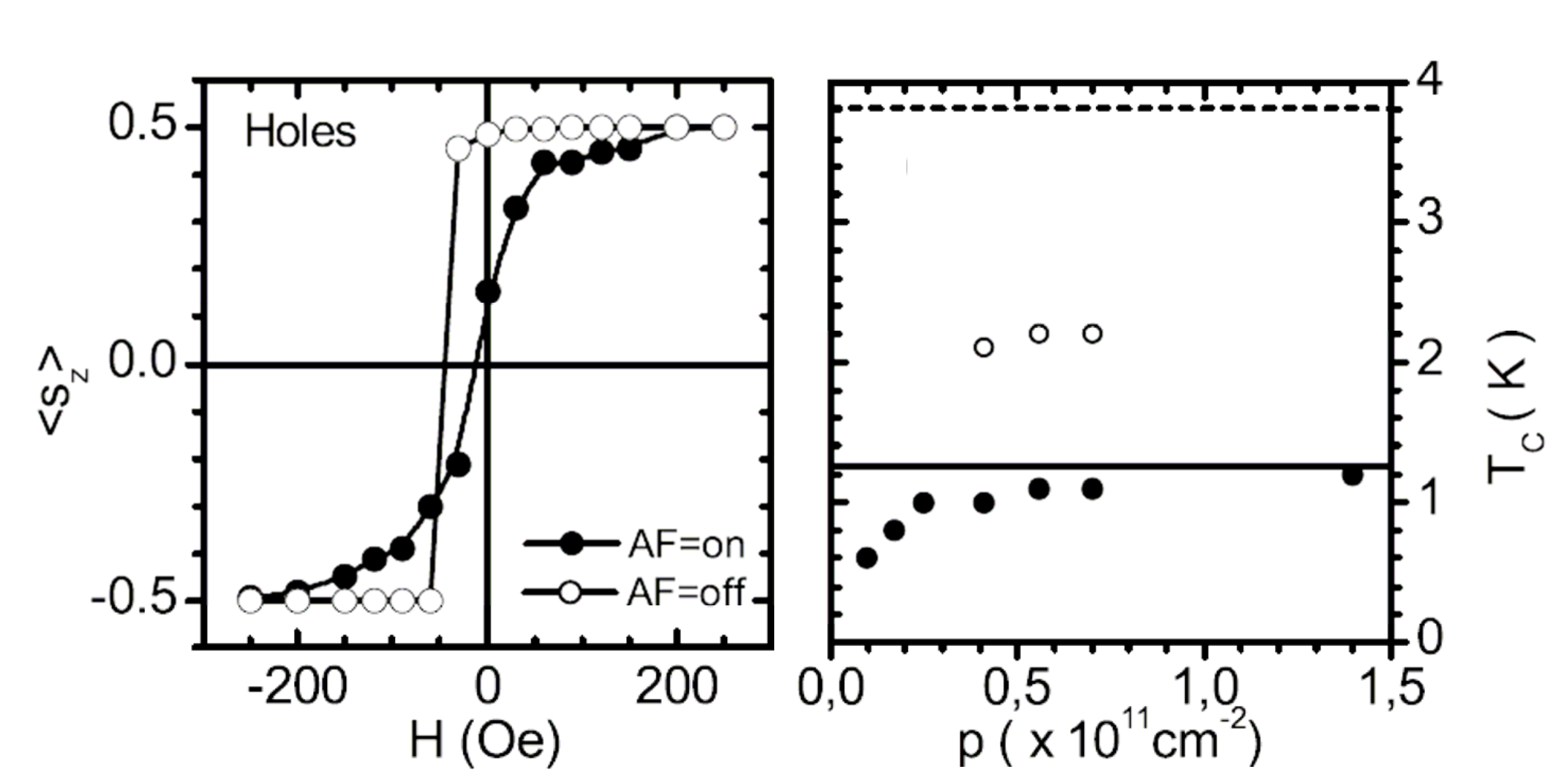}
\caption{Left panel: upper branch of the magnetisation hysteresis loop of hole spins in a Cd$_{0.96}$Mn$_{0.04}$Te quantum well at $T=0.7T_{\mbox{\tiny{C}}}$, where $T_{\mathrm{C}}$ is the Curie temperature, as determined by Monte-Carlo simulations neglecting (open symbols) an taking into account (full symbols) the presence of antiferromagnetic interactions. The sheet hole density is $p=0.41\times  10^{11}$~cm$^{-2}$. Right panel: the simulated dependence of $T_{\mathrm{C}}$ on $p$. Horizontal lines are the mean-field results neglecting (dashed line) and taking into account (solid line) antiferromagnetic interactions (after \cite{Kechrakos:2005_a}).}
\end{figure}

\subsection{Effects of Anderson-Mott localisation}

Similarly to other doped semiconductors, p-type DMS undergo metal-insulator transition (MIT), when an average distance between the carriers becomes two and half times greater than the Bohr radius. The insulator regime can be reached not only by reducing the acceptor density but also by increasing the concentration of compensation donors or by depleting the film of holes  either by electrostatic gates or by charge transfer to surface states or to neighbour undoped layers. In has been found that, in contrast to charge transport characteristics, the Curie temperature, like other thermodynamic properties, does not show up any critical behaviour on crossing the MIT \cite{Matsukura:1998_a,Ferrand:2001_a}.

Two competing models have been put forward in order to explain the existence of ferromagnetic order on the insulator side of the MIT.  According to the magnetic polaron scenario \cite{Durst:2002_a,Kaminski:2002_a} the holes stay localised by the individual parent acceptors, so that their localisation length corresponds to the Bohr radius, usually diminished -- particularly in the strong-coupling regime -- by the hole interaction with the short-range part of the TM potential. In such a case, the ferromagnetic transition can be viewed as the percolation threshold of bound magnetic polarons.

Another scenario was put forward by the present author and co-workers \cite{Dietl:2000_a,Dietl:1997_a,Ferrand:2001_a}.  Within this model, the hole localisation length, which diverges at the MIT,  remains much greater than an average distance between the acceptors for the experimentally important range of the hole densities.  Accordingly, the holes can be regarded as delocalised at the length scale relevant for the coupling between magnetic ions. Hence, the spin-spin exchange interactions are effectively mediated by the itinerant carriers, so that the p-d Zener or Ruderman-Kittel-Kasuya-Yosida (RKKY) model can be applied also on the insulator side of the MIT. At the same time, however, large mesoscopic fluctuations in the local value of the density-of-states are expected near the MIT. As a result, nano-scale phase separation into paramagnetic and ferromagnetic regions takes place below and in the vicinity of the apparent Curie temperature. The paramagnetic phase persists down to the lowest temperatures in the locations that are {\em not} visited by the holes or characterized by a low value of the blocking temperature $T_{\mathrm{B}}$ defined in Introduction. The ferromagnetic order develops in the regions, where the carrier liquid sets long-range ferromagnetic correlation between the randomly distributed TM spins. According to this model, the portion of the material encompassing the ferromagnetic bubbles, and thus the magnitude of the saturated ferromagnetic moment, grows with the net acceptor concentration, extending over the whole sample on the metallic side of the MIT.

It is  still a formidable task, even in non-magnetic semiconductors, to describe quantitatively effects of both disorder and carrier-carrier correlation near the Anderson-Mott transition. However, there is a growing amount of experimental results indicating that the model outlined in the previous paragraph is qualitatively correct. In particular, for samples on the insulator side of MIT, the field dependence of magnetisation shows the presence of superimposed ferromagnetic and paramagnetic contributions in both (Ga,Mn)As \cite{Oiwa:1997_a} and p-(Zn,Mn)Te \cite{Ferrand:2001_a}. Interestingly,  this paramagnetic component is less visible in the anomalous Hall effects as it probes merely the regions visited by the carriers \cite{Ferrand:2001_a}. In these ferromagnetic regions both Mn spins and carriers are spin polarised below $T_{\mathrm{C}}$. Thus, the anomalous Hall resistance being proportional to carrier polarisation shows no paramagnetic component. At the same time, colossal negative magnetoresistance is observed, leading to the field-induced insulator-to-metal transition in samples with the appropriate acceptor densities \cite{Ferrand:2001_a,Katsumoto:1998_a}. The enhanced conductance in the magnetic field can be linked to the ordering of ferromagnetic bubbles and to the alignment of the spins in the paramagnetic regions. Remarkably, the corresponding effects have recently been found in modulation-doped quantum well of (Cd,Mn)Te, where no localisation of carriers by individual ionised impurities and, thus, no formation of bound magnetic polarons is expected \cite{Jaroszynski:2006_a}. The question whether the holes bound by individual acceptors or rather the holes residing in weakly localised states mediate ferromagnetism in DMS on the insulating side of the MIT was also addressed by inelastic neutron scattering in (Zn,Mn)Te:P \cite{Kepa:2003_a}.  In that work, the difference in the the nearest neighbour Mn pairs exchange energy $J_1$ in the presence and in the absence of the holes was determined. The hole-induced contribution to $J_1$ was found to be by a factor of four smaller than that calculated under the assumption that the holes reside on individual acceptors. By contrast, if the hole states are assumed to be metallic-like at length scale of the nearest neighbour distance, the calculated value is smaller than the experimental one by a factor of 1.5, a discrepancy well within combine uncertainties in the input parameters to theory and experimental determination.

\section{Conclusions}

A growing amount of evidences shows that under various conditions the spatial distribution of carries and/or magnetic ions is by no means uniform. The nano-scale phase separation can be driven either by randomness in the carrier and spin subsystems or by limited solubility of transition metals in the host semiconductor, which leads to spinodal decomposition into regions with a small and a large concentration of the magnetic constituent.  Interestingly, by manipulating the charge state of magnetic ions, it becomes possible to control the spinodal decomposition. This constitutes an appealing avenue toward self-organized coherent epitaxy of magnetic nanocrystals over a wide range of their dimensions. It is expected that further works will indicate how to tailor nanocrystal size dispersion and spatial distribution. In this context engineering of local strains by exploiting various combinations of dopants and hosts may turn out to be of relevance. The self-organized growth mode in question \cite{Dietl:2006_a} is rather universal -- it applies to dopants exhibiting a solubility gap, and different charge states that are stable under the growth conditions. The existence of this nano-assembling mechanism, as exemplified here by the case of (Zn,Cr)Te and (Ga,Mn)N co-doped with shallow impurities, explains outstanding properties of a broad class of composite DMS, and offer prospects for exploiting their novel functionalities. In particular, the nanocrystals, rather than the host, are shown to account for ferromagnetic signatures in magnetic and magnetooptical characteristics of these systems.

These findings imply also that today's {\em ab initio} methods of computational materials science, assigning the high-temperature ferromagnetism of, {\em e.~g.}, (Zn,Cr)Te \cite{Bergqvist:2004_a,Fukushima:2004_a} to the uniform diluted alloy, overestimate substantially long-range ferromagnetic correlation, presumably because effects of Mott-Hubbard and Anderson-Mott localisation of paramount importance in the case of the narrow d band are implicitly disregarded in the codes developed so far. It thus appears that delocalised or weakly localised valence band holes are necessary to transmit magnetic information between the diluted spins \cite{Dietl:2000_a}. The accumulated data demonstrate that a number of pertinent properties of spatially uniformed carrier-controlled diluted ferromagnetic semiconductors and their heterostructures can be understood qualitatively, if not quantitatively, by the p-d Zener model, provided that the valence band structure is taken adequately into account \cite{Dietl:2000_a,Jungwirth:2006_a}. A comparison between experimental and theoretical results point clearly to the importance of spin-orbit interactions in the physics of hole-mediated ferromagnetism in semiconductors. These interactions control the magnitude of the Curie temperature, the saturation value of the magnetisation, the anomalous Hall effect as well as the character and magnitude of magnetic and transport anisotropies.

In addition to (Ga,Mn)As, (Zn,Mn)Te:N, and related systems, recent indications of ferromagnetism in p-(Ga,Mn)N \cite{Edmonds:2005_a,Sarigian:2006_a} and p-(Ga,Mn)P \cite{Scarpulla:2005_a}, appear to support the conclusion about the importance of delocalised or weakly localised carriers in mediating efficient spin-spin interactions. However, in those and other experimentally relevant cases, non-uniformity associated with hole localisation, and perhaps with competing ferromagnetic and antiferromagnetic interactions, is seen to affect strongly ferromagnetic properties.

It is still to be experimentally found out whether nitrides, oxides or diamond containing 5\% of randomly distributed magnetic impurities and more than $3\times 10^{20}$ valence band holes per cm$^3$ show ferromagnetic ordering above the room temperature. We may expect a number of important discoveries in the years to come.

\section*{Acknowledgments}

This work was supported in part by NANOSPIN E.~C. project (FP6-2002-IST-015728), by Humboldt Foundation, and carried out in collaboration with M. Sawicki, K. Osuch, H. Przybyli\'nska, M. Kiecana, A. Lipi\'nska, and P. Kossacki in Warsaw, as well as with groups of H. Ohno in Sendai, S. Kuroda in Tsukuba, K. Trohidou in Athens, J. Cibert and D. Ferrand in Grenoble, J. Jaroszy\'nski and  D. Popovi\'c in Tallahassee, L. Molenkamp in W\"urzburg, B. Gallagher in Nottingham, and A. Bonanni in Linz.



\end{document}